\documentclass[twocolumn]{aastex63}

\received{}
\revised{}
\accepted{\today}
\submitjournal{ApJ}

\shorttitle{T CrB}
\shortauthors{Luna et al.}
\graphicspath{{./}{figures/}}
\newcommand{\suzaku}{{\it Suzaku}}
\newcommand{\xmm}{{\it XMM}-Newton}

\newcommand{\ms}{$M_{\odot}$}

\begin{document}

%\title{Increasing pre-outburst activity in T CrB could presage the next nova eruption.}
%\title{Increasing pre-outburst activity in T CrB: forecasting the next nova eruption.}
\title{Increasing activity in T CrB suggests nova eruption is impending.}

\correspondingauthor{Gerardo J. M. Luna}
\email{gjmluna@iafe.uba.ar}

\author{Gerardo J. M. Luna}
\affiliation{CONICET-Universidad de Buenos Aires, Instituto de Astronomía y Física del Espacio (IAFE), Av. Inte. Güiraldes 2620, C1428ZAA, Buenos Aires, Argentina}
\affiliation{Universidad de Buenos Aires, Facultad de Ciencias Exactas y Naturales, Buenos Aires, Argentina.}
\affiliation{Universidad Nacional de Hurlingham, Av. Gdor. Vergara 2222, Villa Tesei, Buenos Aires, Argentina. }

\author{J. L. Sokoloski}
\affiliation{Columbia Astrophysics Lab 550 W120th St., 1027 Pupin Hall, MC 5247 Columbia University, New York, New York 10027, USA}
\affiliation{LSST Corporation, 933 North Cherry Ave, Tucson, AZ 85721.}

\author{Koji Mukai}
\affiliation{CRESST and X-ray Astrophysics Laboratory, NASA Goddard Space Flight Center, Greenbelt, MD 20771, USA}
\affiliation{Department of Physics, University of Maryland, Baltimore County, 1000 Hilltop Circle, Baltimore, MD 21250, USA}

\author{ N. Paul M. Kuin}
\affiliation{Mullard Space Science Laboratory, University College London, Holmbury St Mary, Dorking, Surrey RH5 6NT, UK}



\begin{abstract}

Estimates of the accretion rate in symbiotic recurrent novae (RNe)
often fall short of theoretical expectations by orders of magnitude.
This apparent discrepancy can be resolved if the accumulation of mass by the white dwarf (WD) is highly sporadic, and most observations are performed during low states.
%rather than constant. 
Here we use a reanalysis of archival data from
the Digital Access to a Sky Century @Harvard (DASCH) survey to argue
that the most recent nova eruption in symbiotic RN T CrB, in 1946,
occurred during -- and was therefore triggered by -- a transient
accretion high state.  Based on similarities in the optical light
curve around 1946 and the time of the prior eruption, in 1866, we
suggest that the WD in T CrB accumulates most of the fuel needed to
ignite the thermonuclear runaways (TNRs) during accretion high states.
A natural origin for such 
%occasional accretion high 
states is
dwarf-nova like accretion-disk instabilities, which are expected in
the presumably large 
%accretion 
disks in symbiotic binaries.  The timing of the TNRs in symbiotic RNe could thus be set by the stability
properties of their accretion disks.  T CrB is 
%currently 
in the midst of
an accretion high state like the ones 
%that 
we posit led to the past
two nova eruptions.  Combined with the approach of the
time at which a TNR would be expected based on the 80-year interval
between the prior two novae ($2026 \pm$3), the current accretion high
state increases the likelihood of a TNR occurring in T CrB in the next
few years.

%{\bf not convinced about the abstract }
%T Corona Borealis will probably experience its next nova outburst during the current decade. It presented a particular behavior before its last recorded outburst in 1946: 
%a% re-analysis using 
%ar%chival data from the {\it Digital Access to a Sky Century @Harvard} show that in 1938 it increased its optical brightness by $\Delta$B$\approx$1.5 and this state lasted for about 7 years, fading to the quiescence level for another $\sim$year until nova outburst. 

%The similarities between the brightening that started in early 2014 and is still active and the one from 1938, strongly suggest that this brightening is anticipating the next nova outburst and most likely provide most of the ignition mass necessary to trigger an outburst every $\sim$ 80 years. 
%The next eruption should be expected to happen during the next $\sim$3 years.

\end{abstract}



\section{Introduction} 

\label{sec:intro}

%%% 1st para
T Coronae Borealis (T~CrB) is the nearest symbiotic recurrent nova; a binary system in which a white dwarf (WD) accretes from its red giant companion through a disk. 
%About regularly, 
Twice in the last two centuries, the accreted material ignited on the surface of the WD via runaway thermonuclear fusion reactions and produced 
%what is known as 
a nova eruption. In T~CrB, outbursts were recorded in 1866 and 1946, suggesting a recurrence time of $\sim$80 years. The distance as determined from {\it Gaia} data is 806$^{+33}_{-31}$ pc \citep{2018AJ....156...58B}.
%[Summarize what's know about the distance here -- from Gaia, right?] 
As in the other symbiotic recurrent novae RS~Oph, V745~Sco, and V3890~Sgr, the WD mass is required to be well above 1~\ms\, to generate repeated nova eruptions in less than a century, and so T CrB is a candidate supernova Ia progenitor \citep[e.g.,][]{2000AJ....119.1375F, 1997MNRAS.288.1027S}.  Being the nearest known recurrent nova, we expect the next eruption of T~CrB -- predicted to happen between 2023.6$\pm$1 \citep{tcrb_prediction} and 
%$\sim$ <-Best to not use math symbols in place of words, especially when we don't actually mean order of magnitude.
approximately 2026 -- to shine strongly from $\gamma$-rays to radio wavelengths, providing detailed information about the mass, structure, energetics and perhaps driving mechanism of the outflow.
%an event not to be missed and 

The nova recurrence time is inversely 
%proportional <-Juan, I believe the relationship between the recurrence time and Mdot is not simply \Delta t = constant / M_WD, and same for Mdot.
related to the white dwarf mass and the accretion rate. In the case of a system with the parameters of T~CrB, M$_{WD}$ of 1.2 -- 1.37 M$_{\odot}$ \citep{1998MNRAS.296...77B,2004A&A...415..609S} and recurrence time of about 80 years, theoretical models by \cite{prialnik} predict that an accumulated mass of 10$^{-6}$ \ms~ will be needed for a TNR. Over 80 years that implies an average accretion rate of between 10$^{-8}$ \ms~ yr$^{-1}$ and 10$^{-7}$ \ms~ yr$^{-1}$.

%point out that 
%in order 
%to reach the accumulated mass of about a few 10$^{-6}$ \ms~ needed to have an outburst approximately every 80 years, an average accretion rate during quiescence of between 10$^{-8}$ \ms~ yr$^{-1}$ and 10$^{-7}$ \ms~ yr$^{-1}$ is necessary.  }{\it Paul: I think change this ... predict a an accumulated mass of 10$^{-6}$ \ms~ will be needed for a TNR. Over 80 years that implies an average accretion rate ...}

But measuring the accretion rate in symbiotic binaries is a difficult task. 
% I like that topic sentence, above.  :) (jls)
Although in cataclysmic variables the optical brightness can be used as a proxy for the accretion rate, in symbiotic stars, where the contribution to the optical broadband light of the nebulae and the red giant are not negligible, optical photometry does not provide an actual measurement of the accretion rate.

In the case of T CrB, \cite{1992ApJ...393..289S} and \cite{2004A&A...415..609S} estimated the rate of accretion onto the WD ($\dot{M}_{WD}$)
%[$<-$or whatever symbol we decide to use...]) 
from spectra obtained with the IUE (International Ultraviolet Explorer) satellite between 1978 until 1990. 
%\cite{1992ApJ...393..289S} did not identify periods of high UV flux and estimated an average accretion rate of 9.6$\times$10$^{-9} (d/806 pc)^2$ \ms~yr$^{-1}$. 
Including periods of both high and low UV flux between 1978 and 1990, they found an average accretion rate of 9.6$\times$10$^{-9} (d/806 pc)^2$ \ms~yr$^{-1}$.
%In turn, 
%They obtained an average quiescent accretion rate ($\dot{M}^{ave}_{q}$) 
By modeling the SED in the UV region, \cite{2004A&A...415..609S} obtained a high-state accretion rate (from 1980 to 1988) of 1.1$\times$10$^{-8} (d/806 pc)^2$ \ms~yr$^{-1}$ and a low-state accretion rate (from 1978 to 1980 and 1988-1990) of  1.5$\times$10$^{-9} (d/806 pc)^2$ \ms~yr$^{-1}$.  It should be noted that the high-state identified by \cite{2004A&A...415..609S} did not reach B-magnitudes as bright as the ones discussed here (see Sect. \ref{sec:res}).
%[$<-$ Can we either put "ave" in this definition somehow or an explicit dependence on time, like $\dot{M}_{q}(date)$, given that one of our key points in this paper is that there is no such thing as a simple quiescent accretion rate? (jls)]) 
%of about 5$\times$10$^{-9} (d/806 pc)^2$ \ms~ yr$^{-1}$
%[$<-$ Can we scale this to the Gaia distance and include a $(d/xx)^2$ factor?] 
%by modeling the SED in the UV region.
%the spectra including an accretion disk, a WD and a nebula (check!). 

Additional constraints on $\dot{M}$ onto the WD in T~CrB come from the fact that the boundary layer is typically optically thin, making it a hard X-ray source in quiescence, with the strength of the hard X-ray emission directly related to $\dot{M}$ \citep{2008ASPC..401..342L,2009ApJ...701.1992K}.
%[We should cite other, Luna+ paper here, right?].  
To date, this is a unique feature among 
%the 
all known symbiotic recurrent novae. \suzaku~ observations in 2006 allowed us to measure $\dot{M}_{q}=$ 0.7$\times$10$^{-9} (d/806 pc)^2$ \ms~ yr$^{-1}$ \citep{2019ApJ...880...94L}, where $\dot{M}_{q}$ is the accretion rate measured between nova eruptions. 
%[$<-$ Can we also scale this value to the Gaia distance?]. 
%This accretion rate implies that the boundary layer, the interface between the %Keplerian accretion disk and the WD, is optically thin to its own radiation %\citep{1995ApJ...442..337P,2014A&A...571A..55S}.  <-It's not the accretion rate that tells us the BL is optically thin, right?  We infer that the BL is optically thin from the X-ray spectrum and rapid X-ray flickering, and then because we believe the BL is optically thin, we can infer an Mdot.
Clearly both UV and X-ray measurements indicate that $\dot{M}_{q}$ 
%is 
was low since 1979 until 2006 when compared with the predicted average accretion rate necessary to trigger a nova outburst every 80 years. 

\section{Observations. \label{sec:obs}}

We base the findings described below on publicly available data from two archives -- 
the DASCH \citep[Digital Access to a Sky Century @Harvard;][]{dasch} project to digitize the Harvard Astronomical Photographic Plate collection, which provides a photometric database with a baseline of about 100 years; and $V$- and $B$-band observations from the archive of the American Association of Variable Star Observers (AAVSO).  Details about the DASCH photometric pipeline can be found in \cite{dash_pipeline}. The DASCH database contains unflagged observations of T~CrB from April 1901 through May 1989 with non-uniform sampling.
%which we downloaded and analyzed in the context of the current bright state (see Sect. %\ref{sec:res}).

\section{Results. \label{sec:res}}

\subsection{Optical high state leading up to the 1946 nova eruption}

By querying the DASCH and searching for photometric observations of T CrB previous to the 1946 eruption, we found clear evidence in the $B$-band light curve for an optical bright state that started in 1938 and lasted about 7 years, until about 1945, about one year before the nova eruption, and then continued after the nova event.  Figure 1 shows the DASCH light curve (blue dots).  This light curve confirms the phenomenon mentioned in an abstract by \cite{schaefer2014}, who described finding such a high state associated with both the 1866 and 1946 nova eruptions.  However, the DASCH data revealed that the high state which continued after the eruption for several years had been missed in the previous studies. 
%Previous work had suggest
The visual AAVSO data suggested that a minor precursor event occurred before each eruption that lasted approximately one year.  But our examination of the AAVSO light curve revealed that its coverage around the time of the two novae was too sparse to reveal the full high state that is evident in the DASCH data.
The DASCH light curve also shows that after $\sim$7 years of ``super-active" state, T~CrB faded and almost reached the pre-``super-active" brightness 
%during 
for about one year, after which time the nova 
%outburst 
eruption occurred. 

%The AAVSO $V$ and $B$-band data suggest that about 8 years before the 1946 outburst, %T~CrB brightened for approximately one magnitude during about a one-year.

%Previous to its 1866 and 1946 outbursts, T~CrB presented another distinctive %characteristic 
%%a unique characteristic 
%among the known symbiotic recurrent novae: there was a small optical brightening %($\Delta$B$\approx$1.5, $\Delta$V$\approx$1) that started about 8 years before the %novae eruption \citep{schaefer2014}. 
%%The AAVSO (American Association of Variable Star Observers) data suggested that this %state lasted for about 1 year.

%present in the AAVSO data, in reality

\subsection{Similarity to the current high state}

\begin{center}
\begin{figure}[ht!]
\includegraphics[scale=0.5]{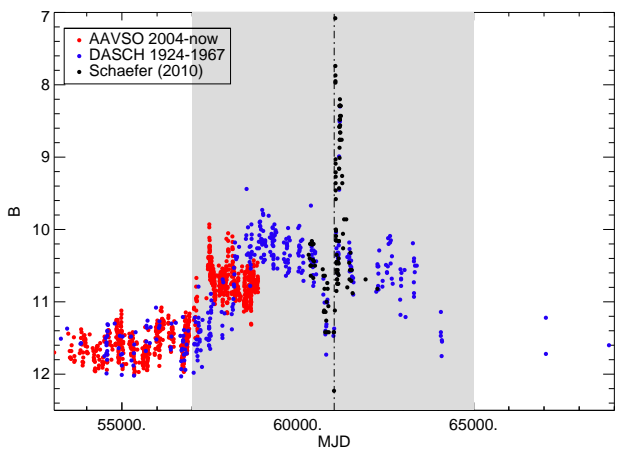}
\caption{DASCH B-magnitude light curve (blue dots) and B-magnitude data (black dots) from Table 12 in \citet{2010ApJS..187..275S}, covering the 1924 to 1967 period, which includes the pre and post-eruption activity and AAVSO B-magnitude light curve covering the 2004-now ``super-active" state (red dots). The DASCH and \citet{2010ApJS..187..275S} light curves were shifted by 80 years (see Sect. \ref{sec:disc}). Vertical dashed line marks the 1946.1 eruption shifted by +80 years. The match of the initial rise between the current and previous brightening is remarkable. The shaded area shows the pre and post-eruption periods during which T~CrB was in a high accretion rate state (see Section \ref{sec:disc}). \label{fig:lc}}
\end{figure}
\end{center}

An on-going optical brightening event that started in early 2014 is extremely similar to the 1938-1945 high state, and the current high state is 
% presumably
driven by an increase in $\dot{M}_{WD}$.   The current optical bright state reached its maximum (B$\sim$10) in April 2016 and has continued since then at an average brightness of B$\sim$10.5 \citep[referred to as a ``super-active" state by][]{munaritcrb}. An \xmm~ observation in January 2017 allowed us to detect a soft X-ray component from the boundary layer, which had become mostly optically thick, and to measure the accretion rate of above 6.6$\times$10$^{-8}$ (d/806 pc)$^{2}$ \ms~ yr$^{-1}$ \citep{tcrb2018}. In March 2018, another \xmm~ observation showed that $\dot{M}$ might have decreased to about 6$\times$10$^{-9}$ (d/806 pc)$^{2}$ \ms~ yr$^{-1}$ \citep{zhekov19}, although a fit to the 2018 \xmm~ spectrum using an alternative model with higher absorbing column suggests the $\dot{M}$ could also have remained high.   In Figure \ref{fig:lc}, we show the DASCH light curve, shifted by 80 years and superimposed upon the AAVSO B-magnitude light curve from the current ``super-active" state. The correspondence between the shape of the initial rise of both events is remarkable. Based on the similar optical behavior in 1938 and 2014, and that both brightenings took place several years before expected nova events, it is likely that previous to the 1946 eruption, the accretion increased to the 10$^{-8}$ \ms~ yr$^{-1}$ level as it did in 2014.  

\section{Discussion. \label{sec:disc}}

The historical light curves show that T~CrB experiences two active states 
%while in quiescence:  
between nova eruptions: one is the so-called ``super-active" state \citep{munaritcrb}, which occurs between $\sim$8 and $\sim$1 year {\it before} the nova eruption and another one, less noticeable, which starts about 200 days {\it after} the nova eruption and lasts for about 8 years.  Although previously
%this has been already 
noted in an abstract by \citet{2014AAS...22320901S}, to our knowledge, the DASCH data show this clearly for the first time. 

We propose that both high states play a significant role in the further development toward the nova eruption. Outside these states, the quiescent accretion rate typically seems to be about 10$^{-9}$ \ms~ yr$^{-1}$; at this level it would not be possible to reach the required ignition mass of a few 10$^{-6}$ \ms~ in 
%$\sim$
approximately 80 years.  If the accretion rate during the high states is about 5$\times$10$^{-9}$ \ms~yr$^{-1}$ to 5$\times$10$^{-8}$ \ms~yr$^{-1}$, then an order of magnitude estimation yields that about 10$^{-6}$ \ms~ could be accreted onto the WD in 20 years, providing a large fraction of the ignition mass. We emphasize however that the nature of the post-eruption high state is unknown, and thus its 
%relation with 
connection to a period of increased accretion rate is only speculative. 

%Thus, most of the ignition mass must be accreted during the $\sim$7 years of the ``super-active'' state and the $\sim$8 years of the post-outburst high state. 
%[Juan, if we consider the duration of the post-TNR optical high state as well, didn't the high state around 1946 last for closer to 15 years?] 

%%%%%{\bf \cite{tcrb_prediction} used the time of rising to the plateau to predict a date for the next TNR. Previous to last outburst in 1946, the rise to the plateau lasted from 1936.0 to 1939.6 while the rise to the current plateau lasted from 2014.2 to 2016.4. By comparing both ``super-active" states, we see that the current state reached the peak earlier than expected if the recurrence time is exactly 80 years.  Moreover, the plateau of the current ``super-active" state is about half {\bf a} magnitude fainter than the 1938-1945 state. We thus hesitate to use the duration of the rising to the plateau as a predictor of the next TNR. However we can use the time of the start of the rising as a rough (?) predictor. From 1936.0 to 2014.2, 78.2 years have passed. Given that the previous nova outburst was in 1946.1, the next of such outbursts could occur on 2024.3.  We note that the fading to the quiescence, pre-outburst state, could have already started. Although the AAVSO light curve does not show a significant decline in the optical brightness, both X-ray softness ratio and H$\alpha$ flux indicate a secular decline in the accretion rate since 2016, but it has not yet reached the quiescence level (Luna et al., {\it in prep.}).  }

By comparing both ``super-active" states, we see that the current state reached the peak earlier than expected if the recurrence time is exactly 80 years \citep[we note however that most recurrent novae do not recur precisely periodically;][]{2010ApJS..187..275S}. Moreover, the plateau of the current ``super-active" state is about half a magnitude fainter than the 1938-1945 state, and thus the fading to the quiescent, pre-eruption state, could have already started. Although the AAVSO light curve does not show a significant decline in the optical brightness, both X-ray softness ratio and H$\alpha$ flux indicate a secular decline in the accretion rate since 2016
%, although it has not yet reached the quiescence level <-JUAN, are you ok with deleting this phrase?
(Luna et al., {\it in prep}.).  We predict that T~CrB is within 3-6 years of its next thermonuclear runaway.  

%We predict that T~CrB is within 3-6 years of its next thermonuclear runaway.

%By comparing both ``super-active" states, we see that the current state reached the peak 
%earlier than expected if the recurrence time is exactly 80 years.  Moreover, the plateau of the current ``super-active" state is about half magnitude fainter than the 1938-1945 state, and thus the fading to the quiescence, pre-outburst state, could have already started. Although the AAVSO light curve does not show a significant decline in the optical brightness, both X-ray softness ratio and H$\alpha$ flux indicate a secular decline in the accretion rate since 2016, although it has not yet reached the quiescence level (Luna et al., {\it in prep}.).  We predict that T~CrB is within 3-6 years of its next thermonuclear runaway. 

%[Do we want to hazard a prediction??]

% Juan, I commented out the text below because I think it weakens our findings.
% This will be a perfect issue to raise in a future observing proposal, but I prefer not %to have it in this paper.
%The origin of this [pre-TNR] fading is unknown and if it repeats after the current "super-active" %state, multiwavelenght observations will be essential to reveal its origin. 

In 
%contrast with 
cataclysmic variables, 
%where 
a single disk instability dwarf nova outburst cannot cause the WD to accumulate enough mass to significantly fuel a nova eruption. For example, \cite{cannizzo93} modeled the disk instability outburst in SS~Cyg and found that about 6$\times$10$^{-10}$ \ms~ yr$^{-1}$ is stored in the disk during quiescence, which is then 
%(at most) 
all or partially dumped into the WD during a long-lasting (about 50 days long) outburst. The aforementioned theoretical models by \cite{prialnik} show that the ignition mass for a nova eruption in a 1.1 \ms ~WD (as in SS~Cyg) is on average 10$^{-6}$ \ms. On the other hand, in symbiotics, where a large and unstable accretion disk can store a significant amount of mass \citep[about $\mathrm{10^{-6}}$ \ms;][]{2008ASPC..401...73W}, the WD has the potential to assemble enough mass after a disk instability to trigger a nova eruption.  

Our results thus suggest a scenario in which the WDs in symbiotic recurrent novae accumulate most of the ignition mass during sporadic high states. Hints of such episodes were already noticed by \cite{2011ApJ...737....7N} in their analysis of quiescent X-ray data of the symbiotic recurrent nova RS~Oph.  Because of their faintness when not experiencing nova eruptions, the data on the other symbiotic recurrent novae (V3890~Sgr, V745~Sco, and perhaps V2487~Oph) are too scarce for us to place constraints on low-amplitude high states like the one for T~CrB we report here.  
With the ignition mass on the WDs in symbiotic recurrent novae most likely to be reached during an accretion high state, states such as the one that T~CrB is currently in then become indicators of an impending nova eruption.

\acknowledgments

We acknowledge with thanks the variable star observations from the AAVSO International Database contributed by observers worldwide and used in this research. The DASCH project at Harvard is grateful for partial support from NSF grants AST-0407380, AST-0909073, and AST-1313370. GJML is a member of the CIC-CONICET (Argentina) and acknowledge support from grants ANPCYT-PICT 0901/2017 and CONICET-NSF International Cooperation Grant 2016. NPMK acknowledges support from the UK Space Agency. JLS acknowledges support from NSF award AST-1616646.

\bibliography{listaref_MASTER}{}
\bibliographystyle{aasjournal}



\end{document}